\documentclass[twocolumn,superscriptaddress,preprintnumbers,amsmath,10pt,aps,prl,nobibnotes]{revtex4} 

\usepackage{algorithm, algorithmic}

\usepackage{amsmath}
\usepackage{amsfonts}
\usepackage{bm}
\usepackage{color}
\usepackage{verbatim}
\usepackage{graphicx}
\usepackage{lipsum}
\usepackage[utf8]{inputenc}
\usepackage{times}
\usepackage{graphicx}
\usepackage{bm}
\usepackage{amssymb,multirow}
\usepackage[dvipsnames]{xcolor}
\usepackage[normalem]{ulem} 
\usepackage[mathlines]{lineno}
\usepackage{bbm}
\usepackage{siunitx}

\def\Veca{\mathbf{a}}
\def\Vecb{\mathbf{b}}

\def\Vect{\mathbf{t}}

\def\Vecw{\mathbf{w}}
\def\Vecx{\mathbf{x}}
\def\Vecy{\mathbf{y}}
\def\Vecz{\mathbf{z}}

\def\VecA{\mathbf{A}}

\def\VecI{\mathbf{I}}

\def\VecP{\mathbf{P}}
\def\VecQ{\mathbf{Q}}

\def\VecS{\mathbf{S}}
\def\VecT{\mathbf{T}}

\def\VecX{\mathbf{X}}

\DeclareMathOperator*{\argmin}{arg\,min} 
\DeclareMathOperator*{\vecop}{vec} 

\def\transpose{^{\!\mathsf{T}}}

\newcommand{\defeq}{\stackrel{\text{\tiny{def}}}{=}}

\begin{document}

\title{Compressively sampling the optical transmission matrix of a multimode fibre}

\author{Shuhui Li}
\email{shli@hust.edu.cn}
\affiliation{Physics and Astronomy, University of Exeter, Exeter, EX4 4QL. UK.}
\affiliation{Wuhan National Laboratory for Optoelectronics, School of Optical and Electronic Information, Huazhong University of Science and Technology, Wuhan 430074, Hubei, China.}
\author{Charles Saunders}
\affiliation{Department of Electrical and Computer Engineering, Boston University, Boston, MA, 02215, USA}
\author{Daniel~J.~Lum}
\affiliation{Department of Physics and Astronomy, University of Rochester, 500 Wilson Blvd, Rochester, NY 14618, USA}
\author{John Murray-Bruce}
\affiliation{Department of Electrical and Computer Engineering, Boston University, Boston, MA, 02215, USA}
\affiliation{Department of Computer Science and Engineering, University of South Florida, Tampa, FL, 33620, USA}
\author{Vivek~K~Goyal}
\affiliation{Department of Electrical and Computer Engineering, Boston University, Boston, MA, 02215, USA}
\author{Tom\'{a}\v{s} \v{C}i\v{z}m\'{a}r}
\affiliation{Leibniz Institute of Photonic Technology, Albert-Einstein-Stra{\ss}e 9, 07745 Jena, Germany}
\affiliation{Institute of Scientific Instruments of CAS, Kr\'{a}lovopolsk\'{a} 147, 612 64, Brno, Czech Republic}
\author{David~B.~Phillips}
\email{d.phillips@exeter.ac.uk}
\affiliation{Physics and Astronomy, University of Exeter, Exeter, EX4 4QL. UK.}

\begin{abstract}
Measurement of the optical transmission matrix (TM) of an opaque material is an advanced form of space-variant aberration correction. Beyond imaging, TM-based methods are emerging in a range of fields including optical communications, optical micro-manipulation, and optical computing. In many cases the TM is very sensitive to perturbations in the configuration of the scattering medium it represents. Therefore applications often require an up-to-the-minute characterisation of the fragile TM, typically entailing hundreds to thousands of probe measurements. In this work we explore how these measurement requirements can be relaxed using the framework of compressive sensing: incorporation of prior information enables accurate estimation from fewer measurements than the dimensionality of the TM we aim to reconstruct. Examples of such priors include knowledge of a memory effect linking input and output fields, an approximate model of the optical system, or a recent but degraded TM measurement. We demonstrate this concept by reconstructing a full-size TM of a multimode fibre supporting 754 modes at compression ratios down to $\sim$5\% with good fidelity. The level of compression achievable is dependent upon the strength of our priors. We show in this case that imaging is still possible using TMs reconstructed at compression ratios down to $\sim$1\% (8 probe measurements). This compressive TM sampling strategy is quite general and may be applied to any form of scattering system about which we have some prior knowledge, including diffusers, thin layers of tissue, fibre optics of any known refractive profile, and reflections from opaque walls. These approaches offer a route to measurement of high-dimensional TMs quickly or with access to limited numbers of measurements.

\end{abstract}
\maketitle

The scattering of light was long thought to be an insurmountable barrier preventing imaging through opaque materials.
However, elastic scattering from static objects is deterministic, and in the last decade it was shown that it is possible to use wavefront shaping with spatial light modulators to characterise and subsequently cancel-out complicated scattering effects~\cite{Vellekoop2007,yaqoob2008optical,Cizmar2010}. Therefore, light that has undergone multiple scattering can be unscrambled to see through opaque media, such as frosted glass~\cite{conkey2012high}, biological tissue~\cite{Papadopoulos:2016aa,yoon2020deep}, or multimode optical fibres (MMFs)~\cite{di2011hologram,choi2012scanner,Cizmar2012}.

Measurement of the Transmission Matrix (TM) of the scattering material in question is a powerful way to achieve this light control capability~\cite{Popoff2010}. The TM can be understood as part of the optical response function of a scatterer: it is a linear operator relating a set of input `probe' fields incident on one side of a scatterer to a new set of output fields leaving on the opposite side. Once the TM is characterised, it encodes how {\it any} linear combination of the probing fields will be scrambled, and more importantly, how to unscramble them again~\cite{Popoff2010a}.
This versatile approach simplifies the task of `undoing' scattering effects: connecting light fields on either side of a scatterer thus circumventing the need to consider the interaction of light with the nano-scale structure of the scatterer itself~\cite{Mosk2012,Rotter2017}.

Beyond imaging, the information-rich nature of the high-dimensional TM is finding applications in a growing number of areas. Examples include the identification of principle modes of MMFs to maximise spatial coherence for high-capacity telecoms applications~\cite{carpenter2015observation}; the optimisation of energy delivery inside scattering materials~\cite{hong2018three,yilmaz2019transverse}; the design of optimised optical trapping fields through randomly scattering systems~\cite{ambichl2017focusing,horodynski2020optimal}; and the creation of new forms of all-optical classical~\cite{matthes2019optical} and quantum information processing~\cite{leedumrongwatthanakun2020programmable}.

The output fields emerging from complicated scattering systems result from the interference of light that has taken many different optical paths through the scatterer. This multi-path interference typically renders high-dimensional TMs extremely sensitive to perturbations in the configuration of the systems they represent. Even recently recorded TM measurements tend to degrade over a period of time (e.g.\ minutes to hours depending on the stability of the scatterer in question and the optical system used to characterise it). In order to maintain a high fidelity, the TM of a scattering system typically needs to be regularly characterised. The number of independent `pixels' in images that can be transmitted through disordered media is conventionally proportional to how many linearly independent probe measurements have been made during TM calibration~-- a number that can easily extend into the thousands. Therefore, establishing new ways to accelerate TM measurement is a useful step towards the deployment of TM-reliant technologies in real-world scenarios.

In this work we explore how the number of probe measurements needed to characterise the TM of a scattering system can be reduced. In many cases, we have advanced knowledge of some general characteristics of the TM we wish to measure. These priors may take different forms, including: knowledge of the existence of a `memory effect' giving characteristic statistical relationships between input and output fields~\cite{Bertolotti2012,Judkewitz:2015aa,li2020guide}; access to a model approximating the optical system~\cite{Ploeschner2015}; or a recent but degraded TM measurement on the same or similar object. Here we provide a guide to the incorporation of these priors into TM reconstruction using the framework of compressive sensing~\cite{candes2008introduction}. We experimentally validate this technique by using it to reconstruct the high-fidelity TM of an MMF supporting 754 spatial modes, using only $\sim$38 measurements ($\sim$5\% of the fibre's mode capacity). Furthermore we show that TMs with sufficient fidelity for imaging can be reconstructed using as few as 8 measurements ($\sim$1\% compression). These methods are universal and may be applied to a range of other scattering systems, including thin layers of tissue, optical diffusers, and scattering from opaque walls.\\

\noindent{\bf Concept}

\noindent The monochromatic $N$-dimensional TM, $\mathbf T \in \mathbb{C}^{N \times N}$, describes how an incident field $\mathbf a \in \mathbb{C}^N$ is transformed via propagation through a scatterer into an output field $\mathbf b \in \mathbb{C}^N$, where $\mathbf b = \mathbf T\mathbf a$. Here $\mathbf a$ and $\mathbf b$ are complex-valued column vectors representing the vectorised (reshaped) 2D input and output fields at a single wavelength and polarisation.

Experimentally, an unknown TM is often measured by injecting a sequence of orthogonal input probe fields, the $n^{\textrm{th}}$ input given by $\mathbf a_n$, and recording how they are transformed, by propagation through the scatterer, into corresponding output field $\mathbf b_n$. The output field is typically measured with a camera, and off-axis digital holography with a coherent reference beam can be used to recover both its amplitude and phase from a single image~\cite{sanchez2014off}. The TM of the scatterer, $\mathbf T$, can then be constructed from these measurements by assigning the $n^{\textrm{th}}$ output field $\mathbf b_n$ to the $n^{\textrm{th}}$ column of $\mathbf T$~\cite{vcivzmar2011shaping}. In this construction, the basis in which $\mathbf T$ is represented is inherited from the bases in which the input and output fields are represented~-- but subsequently we are at liberty to numerically transform its representation into any input and output bases of our choosing. From now on, we refer to this reconstruction technique as {\it column-wise} reconstruction. Evidently, the number of independent measurements, $m$, we need to make should be equal to or greater than the number of orthogonal modes, $N$, we wish to control~-- where here we have defined the recording of an entire output field simultaneously in a single camera image, as an individual `measurement'.

Moving to an under-sampled case, consider the following situation: if we have prior knowledge of a basis in which the TM is {\it perfectly} diagonal, we need only make a single measurement to recover all of the complex amplitudes of the elements on the diagonal. In this case we inject a probe field $\mathbf a_1$ consisting of a known superposition of {\it all} of the modes represented in the diagonal basis, and at the output we measure the transformed field $\mathbf b_1$. Our prior tells us there has been no coupling between modes, and so we can numerically decompose $\mathbf b_1$ into the diagonal basis, and find the complex diagonal elements of the TM by inspecting how the amplitude and phase of each mode has changed compared to the known input. This example illustrates that prior knowledge allows us to recover signals from far fewer measurements than the dimension of the signal. In this case we can make the absolute minimum number of measurements (i.e.\ one) as we have complete knowledge of both the sparsifying basis, and the sparsity pattern (i.e.\ power is only found on the diagonal). Our level of prior knowledge is often much weaker than this example, but the field of compressive sensing~\cite{candes2008introduction}, and more generally the concepts of inference, provide the tools to make the best use of any priors we have to reconstruct high-fidelity TMs with reduced numbers of measurements.

To proceed we note that instead of using the column-wise method, we can construct a linear system of equations to which $\mathbf t$, the vectorised form of $\mathbf T$, is the solution:
\begin{equation}
\label{Eqn:sensing}
\mathbf S \mathbf t = \mathbf y,
\end{equation}
where $\mathbf t \in \mathbb{C}^{N^2}$ is a column vector holding the unknown complex elements of the TM, which may be represented in an arbitrary basis of our choosing. $\mathbf S \in \mathbb{C}^{Nm \times N^2}$ is a `sensing' matrix determined by the set of input modes used to probe the TM, and $\mathbf y \in \mathbb{C}^{Nm}$ is a column vector representing the output measurements. The entries of known matrix $\mathbf S$ and vector $\mathbf y$ depend on our choice of basis representation of $\mathbf t$. See Methods for details of how $\mathbf S$ and $\mathbf y$ are constructed from the set of known input and measured output fields.

If the TM is over-sampled (i.e.\ $m > N$) and $\mathbf S$ is full-rank, then $\mathbf t$ may be found by solving Eqn.~\ref{Eqn:sensing} using standard methods that minimise an error term $\eta$ given by the square of the Euclidean norm of the residual: $\eta = \|\VecS \Vect -  \Vecy \|_2^2$, which accounts for any inconsistencies in Eqn.~\ref{Eqn:sensing} due to noise in the measurements.
If the TM is critically-sampled (i.e.\ $m=N$) and $\mathbf S$ is once again full-rank, then $\mathbf t$ may be found by direct inversion.
However, if the TM is under-sampled (i.e.\ $m < N$), then $\mathbf S$ is rank-deficient and Eqn.~\ref{Eqn:sensing} has an infinite number of possible solutions, only one of which represents the true TM\@. Here our task is to use any prior knowledge of the system we may have to constrain the possible solutions of Eqn.~\ref{Eqn:sensing}, and locate a solution close to the correct one. We note that this prior knowledge could also be used to counteract measurement noise in the over-sampled and critically sampled cases.

\begin{figure*}[t]
\includegraphics[width=17cm]{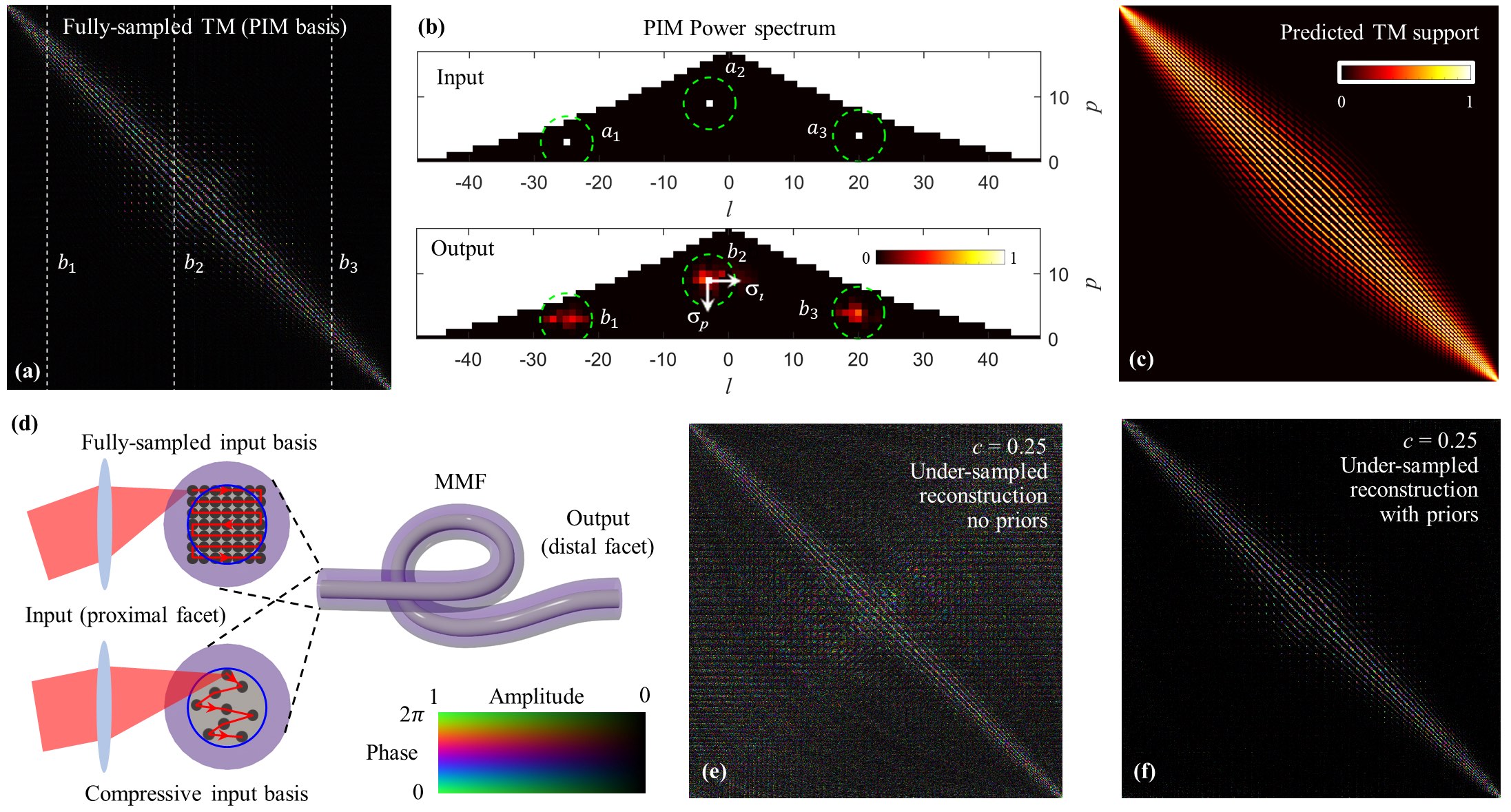}
\caption{{\bf Compressively sampling the TM of an MMF}: {\bf(a)} A fully sampled TM of a 30\,cm segment of step-index MMF supporting 754 modes at a wavelength of 633\,nm (NA = 0.22, core diameter = 50\,$\mu$m). Here represented in the PIM basis, the TM shows strongly diagonal features and a well-defined off-diagonal structure.
{\bf(b)} PIMs are indexed by azimuthal index $\ell$ describing the orbital angular momentum carried by the PIM, and radial index $p$, which is related to the degree of radial momentum carried by the PIM\@. The heat maps in (b) show the power spectra of three input fibre modes ($a_1$, $a_2$, $a_3$) and the three corresponding output fields ($b_1$, $b_2$, $b_3$) represented in the $(\ell,p)$-space of the PIM basis, which also correspond to 3 columns of (a). The colour of pixel $(\ell,p)$ represents the relative power found in mode of index $(\ell,p)$. We see that each input only couples to modes with similar $\ell$ and $p$ indices during propagation through the fibre (also see Movie 1 and SI Fig.~S7). {\bf(c)} A support for the TM can be predicted by estimating the degree of modal coupling $\sigma_\ell$ and $\sigma_p$, indicating areas that are likely to contain little to no power (see Methods for details);
here $\sigma_\ell = 4$ and $\sigma_p = 2$. {\bf(d)} Schematic showing the input modes (a focussed point swept across the input facet) to fully-sample (top) or under-sample (bottom) the TM of the MMF\@. In the over-sampled case, the input focussed beam is scanned across an overlapping Cartesian grid of points. In the under-sampled case, points are addressed in a hyper-uniform pattern. {\bf(e)} An under-sampled TM ($c=0.25$) reconstructed using the column-wise method with no priors, and then transformed into the PIM basis. In this case the TM is evidently reconstructed with low fidelity: it has a lot of off-diagonal power and its normalised correlation with the fully-sampled TM is 0.23\@. The correlation is calculated as the squared modulus of the normalised overlap integral between the complex-valued reconstructed TM and the fully sampled TM.
{\bf(f)} TM reconstruction using the same data as (e) but using FISTA with predicted support. In this case the TM is reconstructed with higher fidelity and the normalised correlation with the fully sampled TM rises to 0.88\@.  Note: (a) and (f) are reproduced in larger scale in Supplementary Information (SI) Figs.~S3 and S4.}
\label{Fig:priors} 
\end{figure*}

A strong prior is the knowledge of a basis in which each input mode does not scatter into many output modes~-- and so the TM is sparse. An even stronger prior is advanced knowledge of {\it which} output modes each input is likely to scatter into. When might we have access to such priors about the TM of a scattering object? There are several situations that provide information of this sort. Firstly, it was recently highlighted that if a scattering system is known to possess a memory effect, then this is equivalent to the knowledge of a basis in which the TM of the scatterer is quasi-diagonal~-- meaning that a significant proportion of the power is found on the main diagonal~\cite{Judkewitz:2015aa,li2020guide}. Secondly, if we have access to a model approximating the optical system in question, then we can use this to find a quasi-diagonalizing basis by simulating the TM and diagonalising it. Thirdly, if we have made a recent but degraded TM measurement on the same scatterer, then this can also be diagonalised to find a sparse basis. We emphasise that the situation we consider in this article is when we have advanced knowledge of the general characteristics of the TM we wish to find, but none of the above alone reveal enough information to build an accurate TM\@~-- so we still need to make {\it some} probe measurements. Our aim is to use the available prior information, along with a small number of new measurements, to reconstruct an accurate TM of the scatterer in question.\\

\vspace{-6mm}
\noindent{\bf Compressively sampling the TM of an MMF}

\noindent We now consider the example of a multimode optical fibre (MMF). Control of light fields through MMFs has attracted growing attention recently as MMF-based micro-endoscopy promises video-rate sub-cellular resolution imaging deep within tissue, at the tip of a needle~\cite{papadopoulos2013high,ohayon2018minimally,Turtaev2018}. MMFs have also been used as mixing elements for classical and quantum optical computing schemes~\cite{matthes2019optical,leedumrongwatthanakun2020programmable}. Modal dispersion means that an image projected onto one end of an MMF is scrambled into a speckle pattern at the other end, and so before an MMF can be deployed as a micro-endoscope, it is necessary to characterise its TM to first understand how to invert this scrambling process~\cite{di2011hologram,choi2012scanner,Cizmar2012}. Unfortunately, any slight bending deformations or temperature fluctuations modify the TM and so quickly degrade imaging performance of current fibre technology~\cite{Flaes2018}. Therefore in the context of emerging MMF-based clinical imaging scenarios, these stability constraints mean that the TM of MMFs may need to be regularly characterised.  

The approximate cylindrical symmetry of an MMF tells us a lot about the structure of the TM in advance of its measurement. Solving the monochromatic wave equation in an idealised straight section of step-index fibre reveals a set of orthogonal circularly polarised Eigenmodes, known as Propagation Invariant Modes (PIMs)~\cite{Ploeschner2015}. The PIMs maintain a constant spatial profile and polarisation on propagation. This means that in the ideal case, power does not couple between these Eigenmodes, and the TM in the PIM basis is unitary and perfectly diagonal. This implies that ideal fibres have a $2\pi$ rotational memory effect~\cite{amitonova2015rotational}, and a quasi-radial memory effect that reaches over the entire output facet~\cite{li2020guide}. Even though real optical fibres differ from this idealised case, Pl{\"o}schner et al.\ recently showed that the TM of a short length of step-index MMF is relatively sparse and strongly diagonal when represented in the PIM basis~\cite{Ploeschner2015}. Detail of the PIMs, and how they are calculated, is given, for example, in refs.~\cite{li2020guide} and~\cite{Ploeschner2015}.

Figure~\ref{Fig:priors}(a) shows an example of an experimentally measured fully-sampled TM of a strand of step-index MMF, represented in the PIM basis. In our experiments we typically find $\sim$10\% of the power is on the main diagonal, and power is concentrated into relatively few elements, meaning the TM is sparse as anticipated by our model. Intriguingly, in this case we are also able to make an estimate of the sparsity pattern, because the remaining power is spread away from the diagonal in a well-structured manner. The root of this structure is revealed by consideration of how the PIMs couple preferentially to others of similar azimuthal ($\ell$) and radial ($p$) mode indices when they are distorted a small amount. For example, the experimentally measured coupling of three input fibre modes that have undergone propagation through a fibre is shown on power spectra plots in Fig.~\ref{Fig:priors}(b), where we see power is only coupled locally in this representation. SI Movie 1 displays the experimentally measured power coupling of every input PIM (also see description in SI Fig.~S7). Evidently this shows that we can directly probe the transformation of multiple input PIMs simultaneously in a single output camera frame, as long as the inputs have well-separated mode indices. Using the example shown in Fig.~\ref{Fig:priors}(b), the transformation experienced by each of the three input modes can be separately measured at the output by transforming the field into the PIM basis, and associating each `island' of power with each individual input PIM\@.

To proceed, we can model the local coupling of PIMs as a 2D Gaussian function with standard deviations $\sigma_\ell$ and $\sigma_p$ describing the degree of power overspill into adjacent modes. This enables prediction of a map capturing the off-diagonal structure we expect to observe in the TM, i.e.\ an estimate of the amplitude of the TM, as shown in Fig.~\ref{Fig:priors}(c). This information can be used as an estimated `support' to guide the TM reconstruction. Prediction of this support is parameterised by estimation of just two numbers ($\sigma_\ell$ and $\sigma_p$). Therefore, for short lengths of MMF (up to tens of centimetres in length~\cite{Ploeschner2015}) of known core diameter and numerical aperture, both the sparsifying PIM basis (i.e.\ the transformation matrix from real-space to the PIM-space) and an estimate of the support are known in advance of any measurements, and can be used in the reconstruction of the TM\@. We also note that by judicious choice of the first few probe measurements, a relatively accurate estimate of $\sigma_\ell$ and $\sigma_p$ can be found~\footnote{Note: By injecting individual PIMs for the first few measurements, the mode coupling of these can be directly measured, which can then be used to estimate $\sigma_\ell$ and $\sigma_p$ to predict the shape of the support.}.

We are now equipped with strong priors about the TM of the MMF in advance of its measurement. So what measurements should we make? As our example at the start of the Concept section illustrated, a good measurement basis is {\it incoherent} with respect to the predicted sparse basis, i.e.\ each of our reduced number of probe measurements should excite many PIMs~\cite{candes2008introduction}. Ideally, all measurements should also be orthogonal to one another to ensure each new measurement yields independent information about the scatterer. To satisfy these requirements, we perform measurements in a basis formed by a single diffraction limited spot that can be focussed onto different places across the core of the input facet of the MMF\@. Each of these foci excite many PIMs and so has a high level of incoherence with the sparse TM basis (see SI Fig.~S1). The spot locations are drawn from a disordered hyper-uniform array which ensures that they do not overlap and so the inputs are orthogonal. An example is shown in Fig.~\ref{Fig:priors}(d), and more detail of how this array is designed is given in Methods. This probing basis also has the advantage of being experimentally straight-forward to accurately create.

To reconstruct the full TM from our under-sampled measurement set, we incorporate our priors by solving the following optimisation problem:
\begin{equation}\label{Eqn:optimise}
\hat{\Vect}  =  \argmin_{\Vect} \tfrac{1}{2}\underbrace{\|\VecS \Vect -  \Vecy \|_2^2}_\text{Data fidelity} + \underbrace{\lambda (\mathbbm{1}-\Vecw)^{\intercal} |\Vect|  }_\text{Sparsity},
\end{equation}
where $\hat{\Vect} \in \mathbb{C}^{N^2}$ is the final solution, and $\Vect \in \mathbb{C}^{N^2}$ is the decision variable, both represented in the sparse PIM basis.
Here $\mathbbm{1} \in \mathbb{C}^{N^2}$ is a column vector of ones and $|\Vect|$ is the magnitude of the complex-valued $\Vect$. Equation~\ref{Eqn:optimise} specifies that the solution should both agree with our under-sampled set of measurements (first term on the right hand side), and also be sparse with low absolute values in regions dictated by our estimated support (second term on the right hand side). We minimise the square of the Euclidean norm in the data fidelity term as we expect the noise to be normally distributed. Column vector $\Vecw \in \mathbb{R}^{N^2}$ is the vectorised predicted support with values between 0 and 1, determined a priori by estimation of $\sigma_\ell$ and $\sigma_p$, as described in Methods. It promotes solutions with magnitudes that adhere more closely to our predicted TM structure. Scalar $\lambda$ is a tuneable parameter that weights the relative importance of the fidelity and sparsity terms (see methods for how this is chosen).

The problem defined in Eqn.~\ref{Eqn:optimise} can be solved using a range of methods. Here we use the Fast Iterative Soft-Thresholding Algorithm (FISTA)~\cite{beck2009fast}, chosen because it is capable of rapidly solving relatively large scale problems with low memory requirements. More detail of how this problem is solved, including pseudo-code, is given in Methods.\\
\begin{figure}[b]
\includegraphics[width=8.5cm]{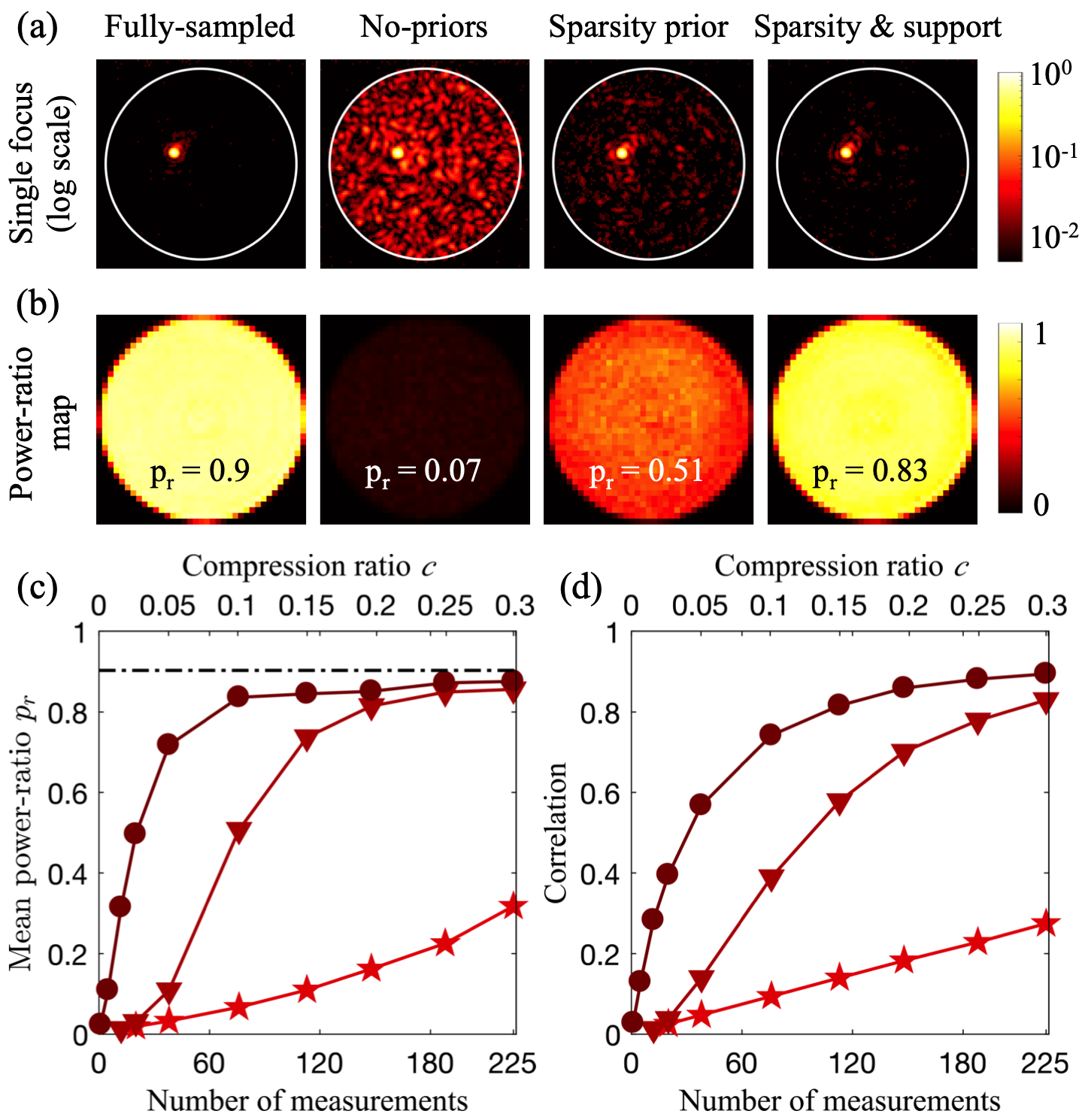}
\caption{{\bf Compressively sampled TM fidelity}: {\bf(a)} Examples of output foci created using TMs reconstructed in different ways at a compression ratio $c = 0.1$. {\bf(b)} Corresponding power ratio maps of spots generated with each reconstruction strategy. {\bf(c)} Graph showing how the mean power-ratio $p_r$ varies as a function of $c$, and the number of measurements $m$, for three different reconstruction strategies incorporating different levels of prior knowledge. The black dashed-dotted line indicates $p_r = 0.9$ of the fully sampled case. {\bf(d)} Graph showing how the level of correlation between under-sampled and fully-sampled TMs varies as a function of $c$. Key for graphs (c) and (d): stars = no priors; triangles = sparsity prior; circles = sparsity and support prior, using the support shown in Fig.~\ref{Fig:priors}(c).}
\label{Fig:compression}
\end{figure}

\vspace{-6mm}
\noindent{\bf Experiment}

\noindent We aim to use an under-sampled set of measurements to reconstruct the TM of a step-index MMF of 30\,cm in length, supporting $N = 754$ propagating spatial modes per polarization at a wavelength of 633\,nm (numerical aperture = 0.22, core diameter = 50\,$\mu$m). These parameters were chosen to reflect those used in prototype MMF-based micro-endoscopes~\cite{Turtaev2018}. The mode capacity of the MMF in this case means that the TM, when represented in the PIM input and output bases, consists of $754^2 = 568516$ complex elements. To reconstruct this TM without exploiting the use of priors requires at least 754 sequentially recorded probe measurements.

Our experimental set-up is shown in SI Fig.~S2\@. The set-up is similar to ref.~\cite{li2020guide}, and is based on a Mach-Zehnder interferometer. In brief: light from the laser source is split into two beam paths. The signal arm of the interferometer contains the MMF under characterisation, along with a Digital Micro-mirror Device (DMD) used to spatially modulate the complex field of the light injected into the MMF~\cite{brown1966complex,lee1979binary,Mitchell2016,Turtaev2017}. The reference arm directs light around the MMF to be used as a coherent reference. The output facet of the MMF is imaged onto a high-speed camera, where it interferes with light from the reference arm, forming an interferogram enabling measurement of the amplitude and phase of the output field in a single camera frame using off-axis digital holography~\cite{sanchez2014off}.

Once the TM is reconstructed, it can be used to create an arbitrary light field $\mathbf d$ at the distal end of the MMF (consisting of any linear combination of the PIMs), by calculating the required proximal field $\mathbf c = \mathbf T^{\dagger}\mathbf d$, where we have assumed the TM is unitary and so $\mathbf T^{-1} = \mathbf T^{\dagger}$. Scanning imaging is achieved by appropriately shaping the input field to sweep a focussed spot over the distal facet~\cite{Cizmar2012}. Reflectance or fluorescence images can be captured by measuring the total reflected/fluorescently excited intensity that is transmitted back to the proximal end, and correlating this signal with each known distal spot location: turning the system into a micro-endoscope~\cite{Turtaev2018}.

To investigate the level of compression experimentally achievable, we probe the TM of the MMF multiple times, in each case reducing the number of measurements $m$, drawn from the hyper-uniform input spot basis (see Methods). The compression ratio $c$ is given by $c=m/N$. For each data-set we compare the performance of three different TM reconstruction algorithms, which incorporate different levels of prior knowledge about the MMF:

\vspace{1mm}
\noindent (i) {\it No priors}~-- column-wise method to reconstruct the TM;

\vspace{1mm}
\noindent (ii): {\it Sparsity prior}~-- FISTA incorporating prior knowledge of the basis in which we estimate the TM to be sparse (i.e.\ PIM basis), but no knowledge of which modes input light is scattered into; i.e.\ no knowledge of the support, and so in this case $\Vecw = 0$ everywhere;

\vspace{1mm}
\noindent (iii): {\it Sparsity prior and estimate of support}~-- FISTA incorporating prior knowledge of both a sparse basis and a TM amplitude support estimate that promotes the diagonal structure, an example of which is shown in Fig.~\ref{Fig:priors}(c). In this case $\Vecw$ is computed from an estimate of $\sigma_\ell$ and $\sigma_p$ (see Methods).

\vspace{1mm}
Figures~\ref{Fig:priors}(e-f) show under-sampled TM reconstructions in the PIM basis ($c\sim0.25$) when using no prior information (Fig.~\ref{Fig:priors}(e)) and when incorporating both sparsity and support priors (Fig.~\ref{Fig:priors}(f)). These can both be compared with the fully sampled TM shown in Fig.~\ref{Fig:priors}(a). Without leveraging priors, the correlation between the under-sampled TM and fully-sampled TM is low. In fact, the correlation is directly proportional to the compression ratio (correlation = 0.23 $\sim c$). Incorporation of priors in the reconstruction significantly boosts the fidelity of the under-sampled TM (correlation of 0.88).

The fidelity of the reconstructed TMs can also be quantified by measuring how well they can be used to generate diffraction limited foci at the output of the fibre. To do this we calculate the mean power-ratio $p_r$, defined as follows: for a Cartesian grid of points across the output facet, we calculate the ratio of power within a small disk centred on the target focus position compared to the total power transmitted through the MMF\@. Figure~\ref{Fig:compression}(a) shows examples of diffraction limited foci generated at the fibre output using TMs reconstructed with the different methods. The mean power-ratio $p_r$ is given by the average power-ratio over all point positions across the core. Figure~\ref{Fig:compression}(b) gives examples of power-ratio maps across the output facet in each case.

 We first benchmark the fidelity of the fully-sampled TM measurement with high signal-to-noise ratio (SNR) by over-sampling the TM with an input basis of a $41\times 41$ Cartesian grid of points (see Fig.~\ref{Fig:priors}(d)). Therefore $c = (41)^2/754 \sim 2.2$\@. Output foci generated using the over-sampled TM incorporating no priors yield an experimental mean power-ratio of $p_r\sim0.9$, demonstrating that the majority of the available power can be focussed to a single point at the output. Several factors contribute to the fact that $p_r<1$ even in the over-sampled case: the accuracy with which the required input field is generated with the DMD; any small drift of the optical system; the hard edge of the disk inside which power is considered in the focus; and low level camera noise~-- which even though low is spread over many pixels compared to the size of the focus. Figure~\ref{Fig:compression}(b),
leftmost panel, shows a map of the power-ratio across the distal facet in this over-sampled case.

Figure~\ref{Fig:compression}(c) shows a graph of mean power-ratio $p_r$ as a function of compression ratio $c$ when applying our different reconstruction strategies. We see that without inclusion of any priors, $p_r$ is once again linearly proportional to $c$, and so for low compression ratios the contrast of spots that can be created on the distal facet is low. This case is equivalent to partial TM measurement, and has been previously considered in, for example, refs.~\cite{Cizmar2012,Turtaev2017,mastiani2019scanning}. By incorporating prior knowledge and reconstructing the TM by solving Eqn.~\ref{Eqn:optimise}, we move to a regime where $p_r>c$. As the compression ratio is reduced, $p_r$ can significantly exceed the $c$ for an under-sampled measurement set. For example, using a sparsity prior alone yields a high-fidelity TM reconstruction, maintaining $p_r>0.8$ down to compression ratios of $c = 0.2$ in this case. This situation is further improved by incorporating the predicted support of the TM, which in this case yields a power-ratio approaching $p_r = 0.9$ when $c = 0.1$ corresponding to only 74 probe measurements. In this case we estimate the level of mode coupling as $\sigma_\ell = 4$ and $\sigma_p = 2$. SI Fig.~S5 shows that at a compression ratio of $c = 0.1$, the reconstruction is relatively robust to variation in these support parameters. Figure~\ref{Fig:compression}(d) shows how the level of correlation between under-sampled and fully-sampled TMs varies as a function of $c$, which also shows a similar trend.

\begin{figure}[t]
\includegraphics[width=8cm]{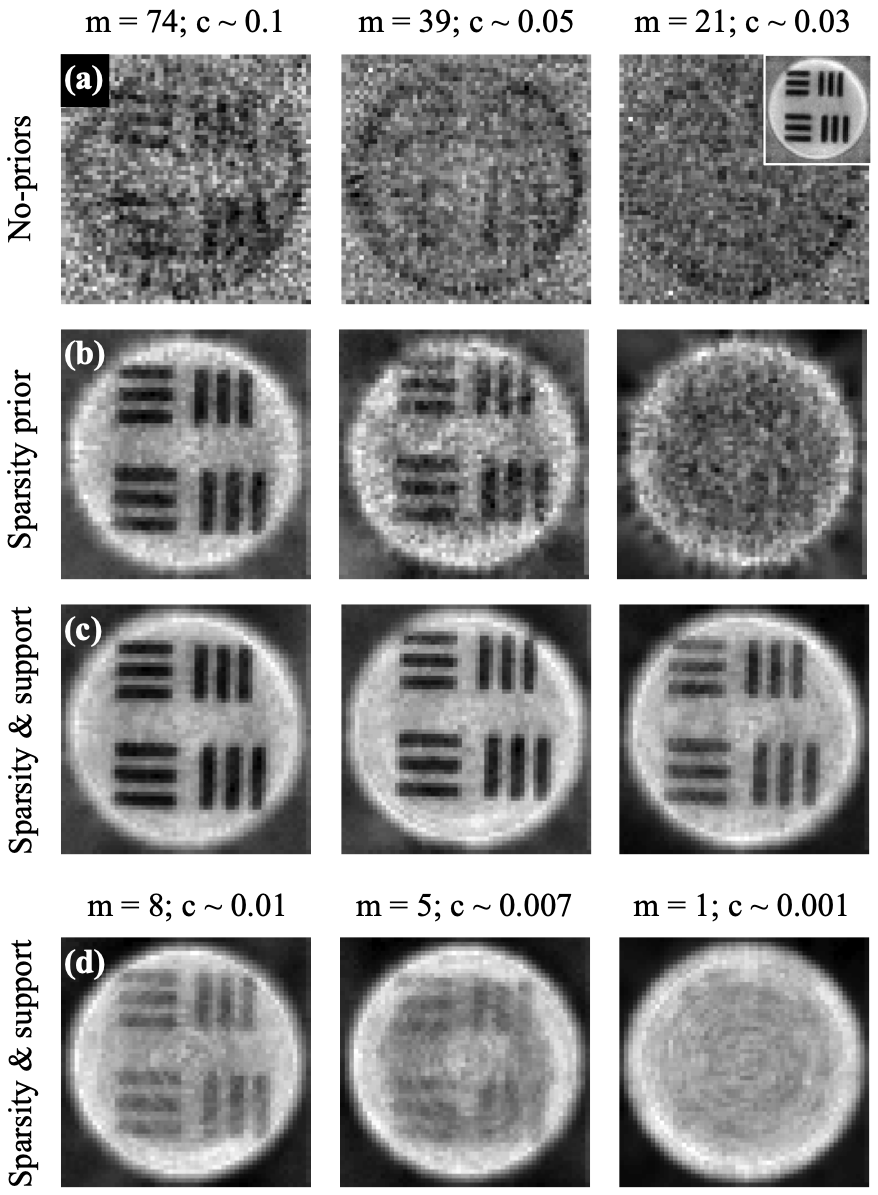}
\caption{{\bf Imaging using compressively recovered TMs}: {\bf(a)} Using a TM reconstructed with no-priors, at progressively reducing levels of sampling from left to right. Imaging is practically impossible at these compression levels. Inset: image captured using a fully sampled TM\@. {\bf(b)} Images from TM reconstructed with a sparsity prior. {\bf(c)} Sparsity and support prior. {\bf(d)} Sparsity and support prior for further reducing levels of sampling. In this case no image is discernible when attempting to infer the TM from a single measurement.}
\label{Fig:imaging}
\end{figure}

We next directly test the imaging performance of the compressively sampled TMs by imaging a resolution target positioned at the output facet of the MMF\@. Figure \ref{Fig:imaging} shows transmission scanning images using foci swept across the output facet that have been generated using TMs reconstructed with the three different strategies. See SI for details. We see that without the use of priors, images of the resolution target are barely discernible at a compression ratio of $c\sim0.1$ (Fig.~\ref{Fig:imaging}(a)). Incorporating a sparsity constraint enables discernible imaging down to $c\sim0.05$ (Fig.~\ref{Fig:imaging}(b)). Inclusion of the support boosts the contrast of imaging at $c\sim0.05$, and enables lower contrast imaging down to $c\sim0.01$, representing just 8 probe measurements (Fig.~\ref{Fig:imaging}(c,d)).

\begin{figure}[t]
\includegraphics[width=8.5cm]{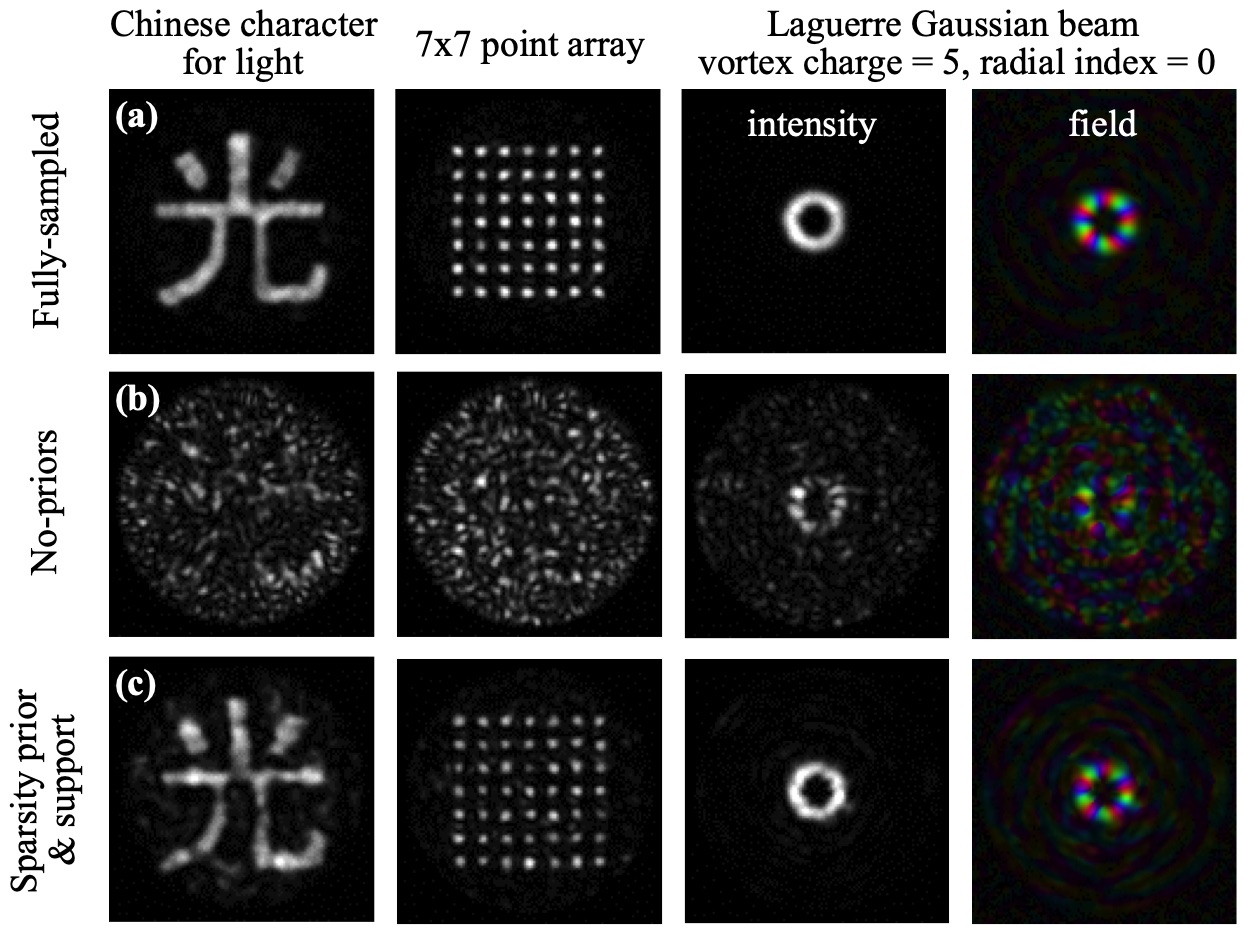}
\caption{{\bf Pattern projection using compressively recovered TMs}: {\bf(a)} Using a fully-sampled TM to project, from left to right: the Chinese character for light, a 7x7 point array, and a Laguerre-Gaussian (LG) beam carrying a vortex charge of $\ell = +5$\@. The rightmost panels show the measured amplitude (brightness) and phase (colour) of the projected LG field reconstructed using digital holography~\cite{sanchez2014off} by processing the interference pattern created by interference with a coherent plane-wave reference beam. Scale-bar is equivalent to that used in Fig.~\ref{Fig:priors}. {\bf(b)} Using an under-sampled TM ($c=0.2$) without priors, it is not possible to project the target patterns. {\bf(c)} Reconstructing the TM using both a sparsity prior and an estimated support, the TM fidelity is high enough to successfully project the target patterns.}
\label{Fig:patterns}
\end{figure}

In addition to scanning imaging, accurate TM reconstructions also enable the projection of arbitrary patterns to the distal facet. The projection of extended patterns is a more challenging test than the creation of focussed spots, as even small inaccuracies in the TM introduce strong speckling effects (i.e.\ extended patterns are more fragile to perturbations). Figure~\ref{Fig:patterns} shows a comparison of the pattern projection capabilities for a TM reconstructed with full sampling (Fig.~\ref{Fig:patterns}(a)), and at a compression ratio of $c\sim0.2$ without priors (Fig.~\ref{Fig:patterns}(b)), and using sparsity priors and support (Fig.~\ref{Fig:patterns}(c)). We test the system by generating the Chinese character for light, a 7$\times$7 array of points, and a Laguerre-Gaussian beam. Evidently at a compression ratio of $c\sim0.2$, it is virtually impossible to project patterns through the fibre without the use of priors in the TM reconstruction.\\

\noindent{\bf Discussion and conclusions}

\noindent In this article we have shown how the framework of compressive sensing can be employed to reduce the number of measurements required to reconstruct high-dimensional optical transmission matrices. Here we have demonstrated this approach to measure the TM of an MMF, but the method is applicable to any scattering system for which we have access to some prior knowledge about the basis in which the TM is likely to be sparse. For example, diffusers, thin layers of biological tissue, and opaque walls all exhibit a tilt-tilt memory effect~\cite{Bertolotti2012,li2020guide} and so have a quasi-diagonal TM in the real-space basis if the input and output planes are placed immediately adjacent to the scattering objects themselves~\footnote{Note: If the planes are placed elsewhere, then as long as the distances from the scatterer to the chosen input and output planes are known, the input and output planes can be digitally transformed to planes adjacent to the object using, for example, the angular spectrum method.}. Therefore, in these cases, compressive TM measurement could be achieved by making an under-sampled set of Fourier basis probe measurements (i.e.\ planes-waves incident from a range of different angles), and then iteratively reconstructing the TM enforcing sparsity in the real-space basis. In diffusive media an estimate of the real-space sparsity pattern could be made, for example, by predicting the degree of lateral modal cross-talk based on the level of diffusion expected through a sample of known transport mean-free-path and thickness. Interestingly, some thin anisotropically scattering samples exhibit quasi-diagonal TMs in both the real-space basis and the momentum-space basis~\cite{Judkewitz:2015aa,Osnabrugge:17}. In this case our algorithm could be extended to enforce sparsity in both of these bases~-- with the higher level of prior knowledge potentially leading to higher compression ratios.

Clearly the compression ratios that may be achieved depend heavily on the strength of prior information available. Using sparsity priors only, the level of compression we can expect is governed by $c\sim s\log(N)$~\cite{candes2008introduction}, where $s$ is the level of sparsity of the TM, i.e.\ the fraction of TM elements that contain appreciable power. As we have shown, the achievable compression level is further improved with additional information about the sparsity pattern.

In our experiment, a key factor in determining accurate estimation of the basis in which the TM is diagonal is alignment of the input to and output from the MMF\@.  This alignment is non-trivial as there are six degrees-of-freedom to consider at each end: position of the objective lens with respect to the fibre in $x$, $y$, and $z$, and tip, tilt and defocus. Previous work showed that a coarse indication of the level of misalignment of these degrees-of-freedom could be extracted from the fully sampled TM of the MMF itself, and hence, digitally corrected; see SI of ref.~\cite{Ploeschner2015}. In this work we found that compressively sampled TMs, when represented in the real-space basis, also provide coarse information on the level of misalignment. Therefore, in order to accurately estimate the PIM basis, we manually align the experimental system (see SI Fig.~S2), and perform an under-sampled set of measurements. We then analyse the raw data to extract coarse estimates of input and output misalignments. These misalignments are then digitally corrected by absorbing them into the real-space to PIM matrices used to transform the TM to the PIM basis for TM reconstruction to commence~\cite{li2020guide,Ploeschner2015}.

The time taken to perform the iterative TM reconstruction is also worth considering. In this work FISTA typically took $\sim$45\,s, however we note that there is scope to significantly reduce this optimisation time to less than a second (see Methods). We also tested the performance of a faster reconstruction method based on Tikhonov regularisation that exhibited lower fidelity but took only 4\,s to complete (see SI Fig.~S6). One area in which we expect compressive sampling of TMs to come into its own, is in situations where measurements cannot be made rapidly, for example when using lower modulation rate phase-only spatial light modulators, or due to low SNR. Compressive sampling also has potential in higher-dimensional cases, such as the measurement of multispectral TMs where the number of measurements can run into the hundreds of thousands~\cite{mounaix2016spatiotemporal,boniface2019rapid,tragaardh2019label,mounaix2019control}.

We note that previous work has also implied the use of compressive TM recovery in some specific cases. For example Gordon et al.\  recently showed that the TM of a fibre bundle can be recovered using fewer output images than the number of fibres, by noting that fibres only couple to their neighbours (i.e.\ the TM is sparse in real-space)~\cite{gordon2019full}. Carpenter et al.\ have previously highlighted that the TM of a graded-index MMF can be approximately represented as a block diagonal structure, which means it is only necessary to measure coupling within the blocks~\cite{carpenter2014110x110}. Our aim here is to highlight that compressive sampling may be applied to reconstruct TMs that are sparse in {\it any} known basis, with a number of camera frames that is lower than mode capacity of the system. We also show how knowledge of the sparsity pattern can be leveraged, and present the first experimental demonstration (to our knowledge) of compressively sampling the TM of an MMF using this technique.

Finally, we highlight that the concept of compressive TM reconstruction may be interpreted as constrained phase-retrieval in a large number of dimensions. In phase-retrieval, the objective is typically to estimate unknown phase components of a complex field with access only to the intensity of the field and some constraints~\cite{fienup1982phase,dremeau2015reference}. Here we have access to both the intensity and phase of an under-sampled set of measurements that are linked through a high-dimensional linear system of equations (i.e.\ the TM)~-- and the iterative approach we have used to solve this problem is similar to those used in phase retrieval problems~\cite{howland2014compressive,mirhosseini2014compressive}. More broadly, the concepts of compressive sensing have been combined with the high-dimensional transformations enacted by scattering systems in several other ways in the past. Most notably compressive sensing has been applied to reduce the number of measurements required to recover images through scattering systems, by drawing on priors about the form of the images themselves~\cite{Liutkus2014,amitonova2018compressive,caravaca2019hybrid}. Our work complements these previous studies, by highlighting that it is also possible to draw on priors relating to the {\it scatterer itself} during the calibration phase. In the future, we hope that compressive TM reconstruction concepts can be coupled with ultra-fast modulators currently under development~\cite{feldkhun2019focusing,tzang2019wavefront}, unlocking the potential to characterise and image through even dynamically changing scattering systems as efficiently as possible.


\section{Acknowledgements}
DBP and SL thank Sergey Turtaev and Ivo Leite for advice on setting up the experiment. SL acknowledges support from the National Natural Science Foundation of China under Grant no.\ 61705073\@. CS, JMB, and VKG acknowledge support from the United States National Science Foundation under Grant no.~1815896\@. TC thanks the European Regional Development Fund (CZ.\ 02.1.01/0.0/0.0/15\_003/0000476) and the European Research Council (724530) for financial support.
DBP thanks the Royal Academy of Engineering and the European Research Council (804626) for financial support.

\section{Author contributions}
DBP, DJL and TC conceived the idea for the project. DBP supervised the project. DBP validated the concept in simulation and developed the Tikhonov regularisation-based reconstruction algorithm. SL built the optical setup and performed all experiments and analysis, with support from DBP and TC\@. CS, JM-B, and VKG developed the FISTA-based reconstruction algorithm. DBP and SL wrote the manuscript with editorial input from all other authors.

\section{Methods}

\noindent{\bf Constructing the sensing matrix}:
Consider an MMF with a mode capacity of $N$ modes per polarisation, with unknown TM $\VecT\in \mathbb{C}^{N\times N}$, represented in the PIM basis. In our work the PIM basis is the natural choice as it is the basis in which the sparsity priors are enforced.

To fully sample the $N\times N$-element TM, we inject $N$ orthogonal probe modes, $\Veca_{1}, \Veca_{2}, \ldots, \Veca_{N} \in \mathbb{C}^{N}$, also expressed in the PIM basis. The transformation of these inputs by the MMF produces the following outputs:
\begin{align*}
\mathbf{b}_1  &= \VecT\mathbf{a}_1,\\
\mathbf{b}_2  &=  \VecT\mathbf{a}_2,\\
&~~\vdots \\
\mathbf{b}_N  &=  \VecT\mathbf{a}_N.
\end{align*}
Horizontally concatenating the corresponding sides of each of the above equations, and taking the transpose of the resulting matrix, we obtain the matrix product:
\begin{align}
    \begin{bmatrix}\Vecb_1, \, \Vecb_2,\ldots,\,\Vecb_N \end{bmatrix}\transpose
    &= \begin{bmatrix}\VecT\Veca_1, \, \VecT\Veca_2,\ldots,\, \VecT\Veca_N \end{bmatrix}\transpose \nonumber \\
    &= \left( \VecT\begin{bmatrix} \Veca_1, \,\Veca_2,\ldots,\, \Veca_N \end{bmatrix}\right)\transpose \nonumber\\
    &= \begin{bmatrix} \Veca_1, \,\Veca_2,\ldots,\, \Veca_N \end{bmatrix}\transpose \VecT\transpose\nonumber\\
    &=  \VecA{}\VecT \transpose,
    \label{eq:LinSys_matrixform}
\end{align}
where $(\cdot)\transpose$ denotes the matrix transpose operator.
Vectorising both sides of the above equation gives an equivalent matrix-vector form:
\begin{align}
    \vecop \left(\begin{bmatrix}\Vecb_1, \, \Vecb_2,\ldots,\,\Vecb_N \end{bmatrix}\transpose\right)
    &= \vecop \left( \VecA{}\VecT \transpose \right)\nonumber\\
    &= \left( \VecI_{N} \otimes \VecA\right)  \vecop \left( \VecT\transpose \right),
    \label{eq:LinSys_matrixvectorform}
\end{align}
where $\vecop(\cdot)$ denotes the vectorisation operator,
$\VecI_{N}$ is the $N\times N$-identity matrix,
and the symbol $\otimes$ denotes the Kronecker matrix product between two matrices. The last equality in Eqn.~\ref{eq:LinSys_matrixvectorform} follows from the vectorization-Kronecker product identity: $\vecop(\VecP \VecX \VecQ) = (\VecQ\transpose \otimes \VecP)\vecop(\VecX)$.
Finally, by letting $\Vecy \defeq \vecop \left(\begin{bmatrix}\Vecb_1, \, \Vecb_2,\ldots,\,\Vecb_N \end{bmatrix}\transpose\right)$, $\VecS \defeq \left(\VecI_N \otimes \VecA \right)$ and $\Vect \defeq \vecop \left(\VecT \transpose\right)$
we obtain the desired form given in Eqn.~\ref{Eqn:sensing}.

Since $\VecS = (\VecI_N \otimes A)$, it is an $N^2 \times N^2$-block diagonal matrix, with $\VecA$ repeated along the diagonal. Each row of $\VecA$ is a single probe mode, here expressed in the PIM basis. Therefore reducing the number of measurements is equivalent to reducing the number of rows in $\mathbf A$, and so also consequently reducing the number of rows in $\mathbf S$, and elements in $\mathbf y$. When the TM is under-sampled, $\mathbf S$ has fewer rows than columns.
Therefore, application of priors within the framework of compressive sensing is necessary to solve Eqn.~\ref{Eqn:sensing}. We also note that the block-diagonal nature of $\mathbf S$ means that the reconstruction of each column of the TM can be carried out independently and in parallel, facilitating rapid TM reconstruction if necessary.\\

\noindent{\bf Estimating the support}:
To construct the $n^{\mathrm{th}}$ column of the predicted TM support, we numerically define a discretised 2D Gaussian function $f$ in $(\ell,p)$-space within the bounds of the power spectra grid representing the indices of the allowed PIMs (see Fig.~\ref{Fig:compression}(b)):
\begin{equation}
f(\ell,p) = \exp\left[-\frac{(\ell-\ell_0)^2}{2\sigma_\ell^2} - \frac{(p-p_0)^2}{2\sigma_p^2}\right],
\end{equation}
centred on the $n^{\mathrm{th}}$ PIM corresponding to indices $\ell_0$ and $p_0$. Here $\sigma_\ell$ and $\sigma_p$ are standard deviations representing our estimated level of coupling. This 2D function is then reshaped into a column vector to form the $n^{\mathrm{th}}$ column of the predicted TM support. This process is repeated with the centre of the Gaussian function moved over each PIM index, to build up the entire predicted support, such as that shown in Fig.~\ref{Fig:priors}. The ordering of the PIMs of indices $\ell$ and $p$ into a 1D list is arbitrary but must be self consistent. Here we follow the ordering used in refs.~\cite{li2020guide} and~\cite{Ploeschner2015}. Finally the predicted support is vectorised to generate $\Vecw$ that is used in Eqn.\ref{Eqn:optimise}.\\

\noindent{\bf Design of the hyper-uniform input basis}:
In order to ensure the input facet is approximately evenly sampled, the locations of $m$ points are selected by creating a hyper-uniform array. We first randomly distribute the $m$ points across a disk representing the core of the fibre. To prevent the clustering that naturally occurs when the locations are randomly chosen, we iteratively update the position of the points to evenly spread them across the core. This is achieved by defining a repulsive `force' acting along the line that joins two points, the magnitude of which is inversely proportional to the distance between the points. The total resulting force vector acting on an individual point is the vector sum of the repulsive forces from all nearby points. On each iteration we move each point in the direction of the total force vector acting on it. The size of the movement is a small distance (on the order of a hundredth of the core radius) proportional to the magnitude of the total force acting on each point. An additional force pointing radially inward is applied to points near the edge of the core to prevent the points repelling each other beyond the radius of the core. The position of the points are updated until no appreciable changes are observed. The resulting set of points then specify the locations of foci of under-sampled input probe measurements.\\

\noindent{\bf Solving the optimisation problem}:
Algorithm~\ref{alg:fista} describes the Fast Iterative Soft-Thresholding steps used to solve the problem in Eqn.~\ref{Eqn:optimise}. This is known also as an \emph{accelerated proximal gradient descent} algorithm.

 \begin{algorithm}[H]
 \caption{FISTA algorithm to solve Eqn.~\ref{Eqn:optimise}}
  \label{alg:fista}
 \begin{algorithmic}[1]
 \renewcommand{\algorithmicrequire}{\textbf{Input:}}
 \renewcommand{\algorithmicensure}{\textbf{Output:}}
 \renewcommand{\algorithmicfor}{\textbf{while}}
 \REQUIRE Initial estimate $\Vect^0$ \\ Measurement vector $\Vecy$ \\ Sensing matrix $\VecS$\\ Regularization strength $\lambda \geq 0$\\  Estimated support vector $\Vecw$\\  Step size $\alpha < 1/L$ ($L$ is Lipschitz constant of the gradient of the cost function).
 \ENSURE  $\hat{\Vect}$
    \STATE $\mu = 1$, $\Vecx^0 = \Vect^0$
  \FOR {not converged}
  \STATE $\Vecx^{k+1} = P_{\alpha \lambda \Vecw}(\Vect^k - \alpha( \underbrace{\VecS \transpose (\VecS \Vect^{k} - \Vecy)}_{\text{Data fidelity gradient}})) $
  \STATE $\mu^{k+1} = \frac{1}{2}(1 + \sqrt{4(\mu^k)^2 + 1})$
  \STATE $\Vect^{k+1} = \Vecx^{k+1} + (\mu^k - 1)/\mu^{k+1} (\Vecx^{k+1} - \Vecx^k)$ 
  \ENDFOR
 \RETURN $\hat{\Vect} = \Vect^{k+1} $ 
 \end{algorithmic} 
 where $P_{\bm{\tau}}(\Vecz)_i = \text{max}(0, 1 - \bm{\tau}_i/|\Vecz_i|)\Vecz_i$ gives the solution to the proximal operator for the sparsity regularization term, where ${\bm{\tau}} = \alpha\lambda\Vecw$.
 \end{algorithm}

We implemented Algorithm 1 in MATLAB. $\lambda$ was manually tuned once by testing the reconstruction performance for a range of choices for $\lambda$, before choosing $\lambda = 0.25$. This value of $\lambda$ was used for all reconstructions, irrespective of the compression ratio. Calculation of the Lipschitz constant used as a bound for the choice of the step size involves computing the singular value decomposition of the sensing matrix, which took $\sim$35\,s at the outset. Alternatively, a simple back-tracking scheme could be used to perform automatic step size selection. $\Vect^0$ was initialised using the solution obtained from the column-wise method, for example see Fig.~\ref{Fig:priors}(e). Running on a laptop with a quad core (8 threads) Intel i7-8565U CPU, with 8\,GB of RAM, it typically took $\sim$45\,s to solve Eqn.~\ref{Eqn:sensing} with a compression ratio of $c\sim0.15$ at a fixed step-size.

 In this case the reconstruction time is longer than the time taken to record the fully-sampled TM using a fast DMD and high-speed camera ($\sim$10\,s, excluding pattern loading time). However, we note that this reconstruction time could be significantly reduced by specifically tailoring the optimisation algorithm to take advantage of the structure of the sampling matrix $\VecS$. Here we treated Eqn.~\ref{Eqn:optimise} as a single sparse matrix equation, but the block-diagonal structure of $\VecS$ means that Eqn.~\ref{Eqn:optimise} is separable into, in this case, $N = 754$ smaller equations that can in principle be solved in parallel to recover each column of the TM independently. We estimate this would reduce the reconstruction time to less than a second.

\onecolumngrid

\vspace{20cm}

\section{Supplementary Information}

\noindent {\bf S1 Transformation matrix from spots to PIMs.\\
S2 Schematic of the experimental set-up.\\
S3 Fully-sampled TM in PIM basis reconstructed using the column-wise method.\\
S4 Under-sampled TM in PIM basis reconstructed using FISTA with sparsity and support priors.\\
S5 Robustness of reconstruction to inaccurate estimation of $\sigma_\ell$ and $\sigma_p$.\\
S6 Tikhonov regularisation results.\\
S7 Movie 1: Experimentally measured coupling of PIMs.
}\\

\renewcommand{\thefigure}{S1}
\begin{figure*}[h]
\includegraphics[width=16cm]{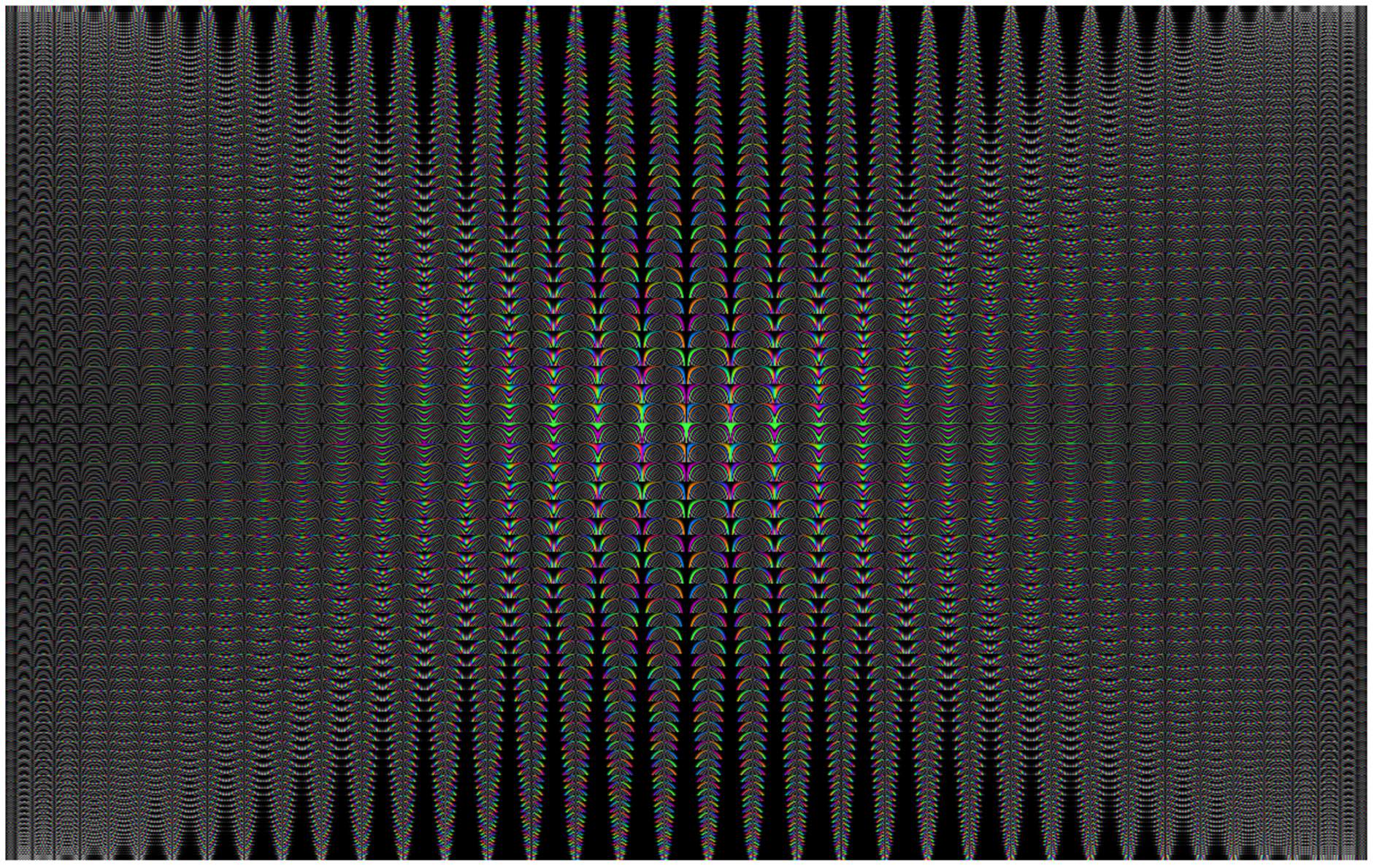}
\caption{{\bf Transformation matrix from spots to PIMs}: Here columns represent spot locations, and rows represent PIMs. Therefore the pixel at row $i$, col.~$j$ represents the amplitude (brightness) and phase (colour) of the overlap between a spot focussed at location $j$ on the input facet, with PIM $i$. We see that the majority of spot locations excite many PIMs~-- showing that the spot input basis is relatively incoherent with the PIM basis as required in our compressive sampling protocol.}
\label{Fig:bigTMunder}
\end{figure*}

\renewcommand{\thefigure}{S2}
\begin{figure*}[h]
\includegraphics[width=16cm]{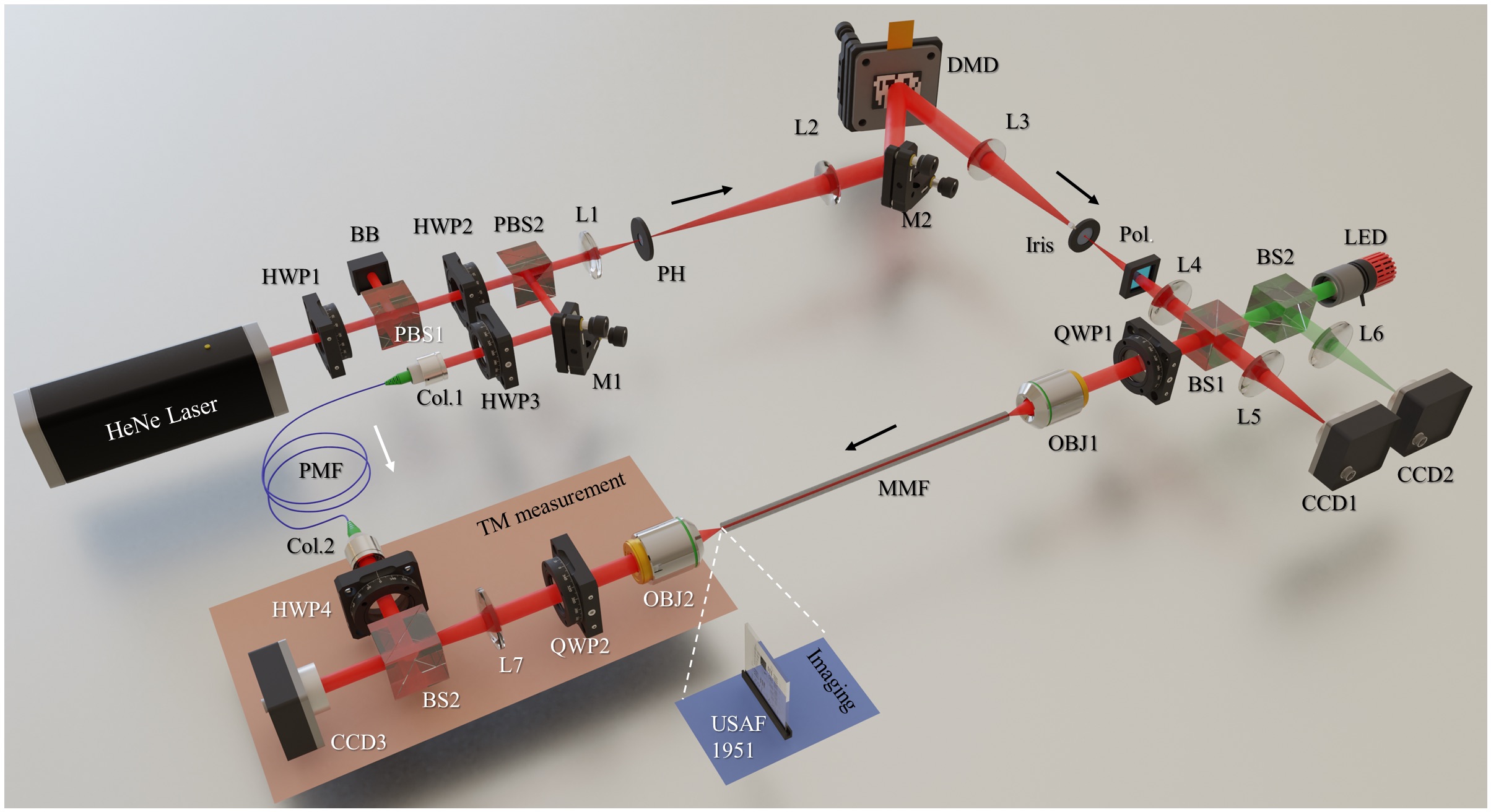}
\caption{{\bf Schematic of the experimental set-up}: The optical system is based on a Mach-Zehnder interferometer. A laser beam is generated by a 1\,mW Helium-Neon laser source operating at a wavelength of 633\,nm. The beam is split into target and reference arms using a polarising beamsplitter (PBS2). In the target arm, the beam is spatially filtered and expanded to fill a digital micro-mirror device (DMD) (ViALUX V-7001). The first diffraction order of the DMD is selected by an iris which blocks all other diffraction orders. The DMD chip plane is imaged onto the pupil of an objective lens (OBJ1, 20X magnification). The input facet of the MMF is placed at the front focal plane of the objective lens. The MMF output facet is imaged onto a high-speed camera (CCD3, Basler Pilot GigE, resolution 648x488), where it interferes with the coherent reference beam. The plane wave of the reference beam arrives at the camera at a small tilt angle with respect to the camera chip normal, enabling single-shot digital holography to reconstruct the intensity and phase of the target field. CCD1 and CCD2 are alignment cameras (also Basler Pilot GigE). They are in the image plane of the proximal facet of the MMF\@. CCD1 images the incident laser beam,  enabling aberration correction of the part of the optical set-up before the MMF if necessary, using, for example, the methods described in ref.~\cite{Cizmar2010}. This correction need only take place once and is unchanged regardless of the test scattering object that is placed in the TM measurement system. A red LED illuminator is used to illuminate the proximal facet of the fibre to aid alignment. Transmission imaging was achieved by scanning a focussed beam over a transmissive resolution target placed $\sim40\mu$m from the distal facet of the fibre. The TM under question was used to calculate the input field required to generate a focussed spot on the output distal facet of the MMF\@. This was then refocussed from the distal fibre facet to the plane of the resolution target by adding a quadratic Fresnel lens phase function to the hologram displayed on the DMD, as described in ref.~\cite{Cizmar2012}. To reconstruct an image, the total transmitted intensity arriving at CCD3 was recorded for each spot location at the distal facet. Reflection imaging is also possible (as would be necessary for the MMF to be deployed in a real application), but in our case the low power of the laser used for the experiment, coupled with low collection efficiency for MMFs meant the returning signals were small, which introduced additional noise and so we did not use reflection images to test the performance of the reconstructed TMs.}
\label{Fig:exp}
\end{figure*}

\renewcommand{\thefigure}{S3}
\begin{figure*}[p]
\includegraphics[width=10cm]{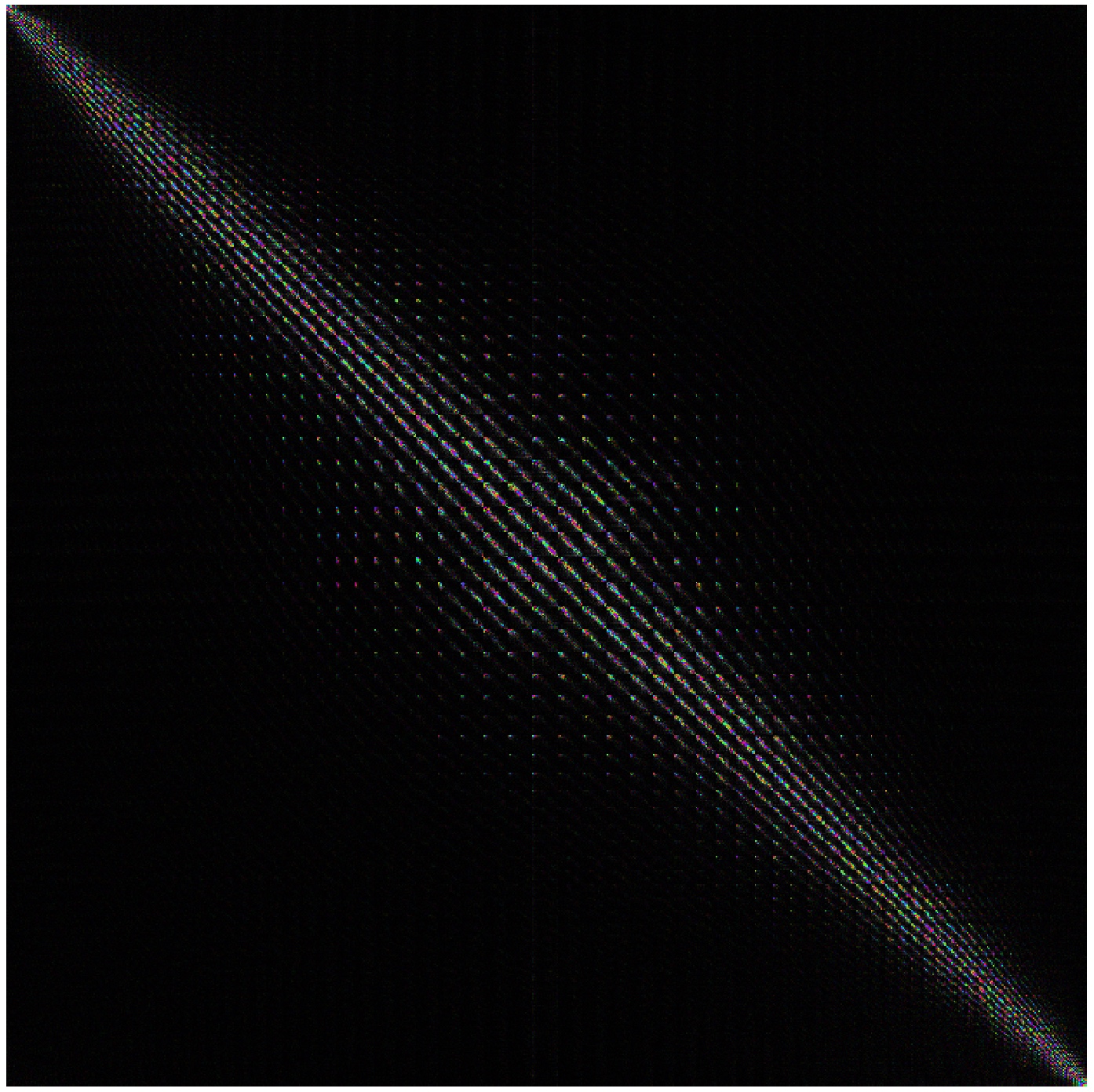}
\caption{{\bf Fully-sampled TM in PIM basis reconstructed using the column-wise method}: This is a larger scale reproduction of Fig.~1(a). The TM is over-sampled using $c = 2.2$ as explained in the main text.}
\label{Fig:bigTMfull}
\end{figure*}

\renewcommand{\thefigure}{S4}
\begin{figure*}[p]
\includegraphics[width=10cm]{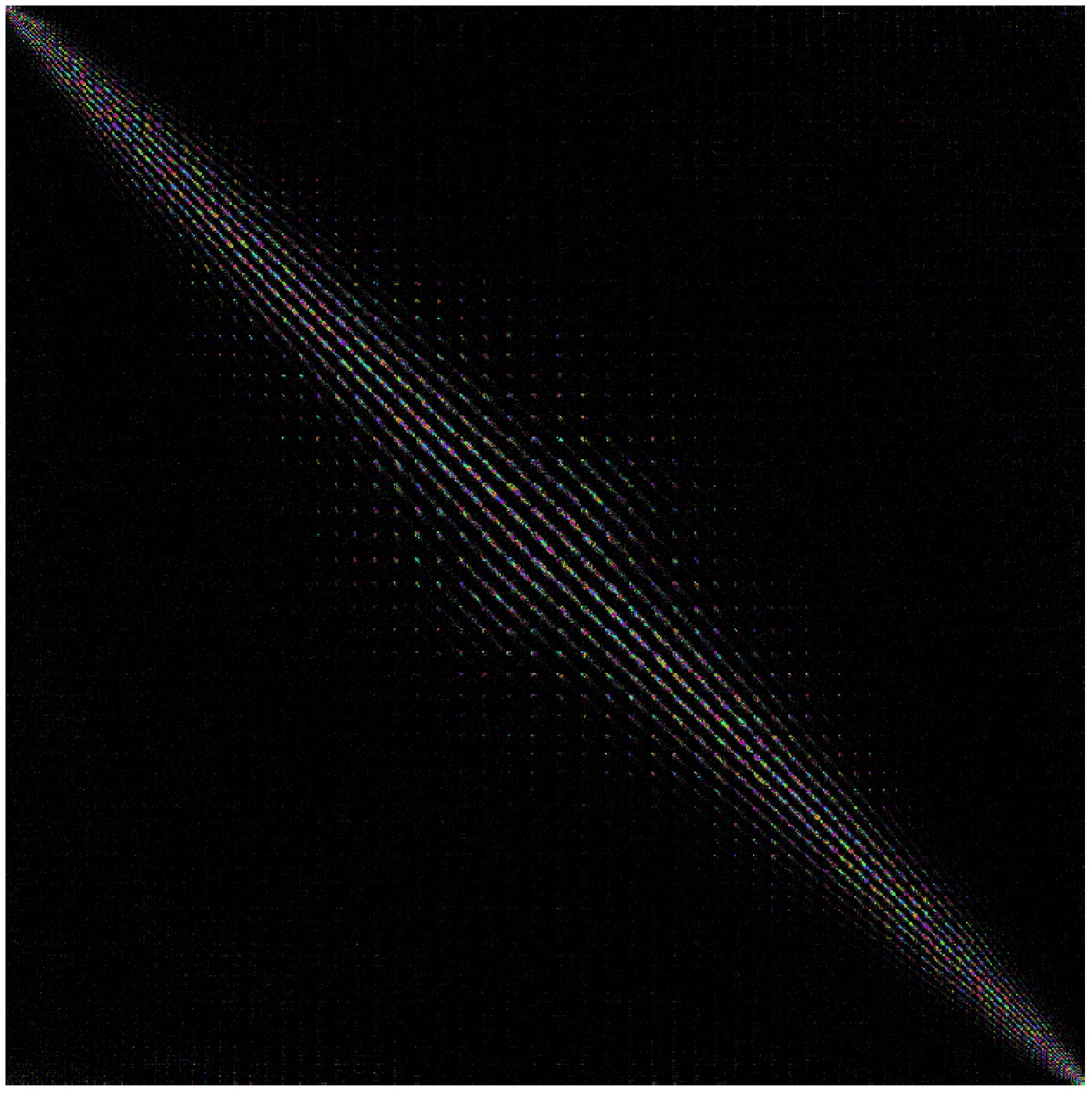}
\caption{{\bf Under-sampled TM in PIM basis reconstructed using FISTA with sparsity and support priors}: This is a larger scale reproduction of Fig.~1(f). The TM is under-sampled using $c = 0.25$. The correlation with the fully sampled TM is 0.88.}
\label{Fig:bigTMunder}
\end{figure*}

\renewcommand{\thefigure}{S5}
\begin{figure*}[h]
\includegraphics[width=15cm]{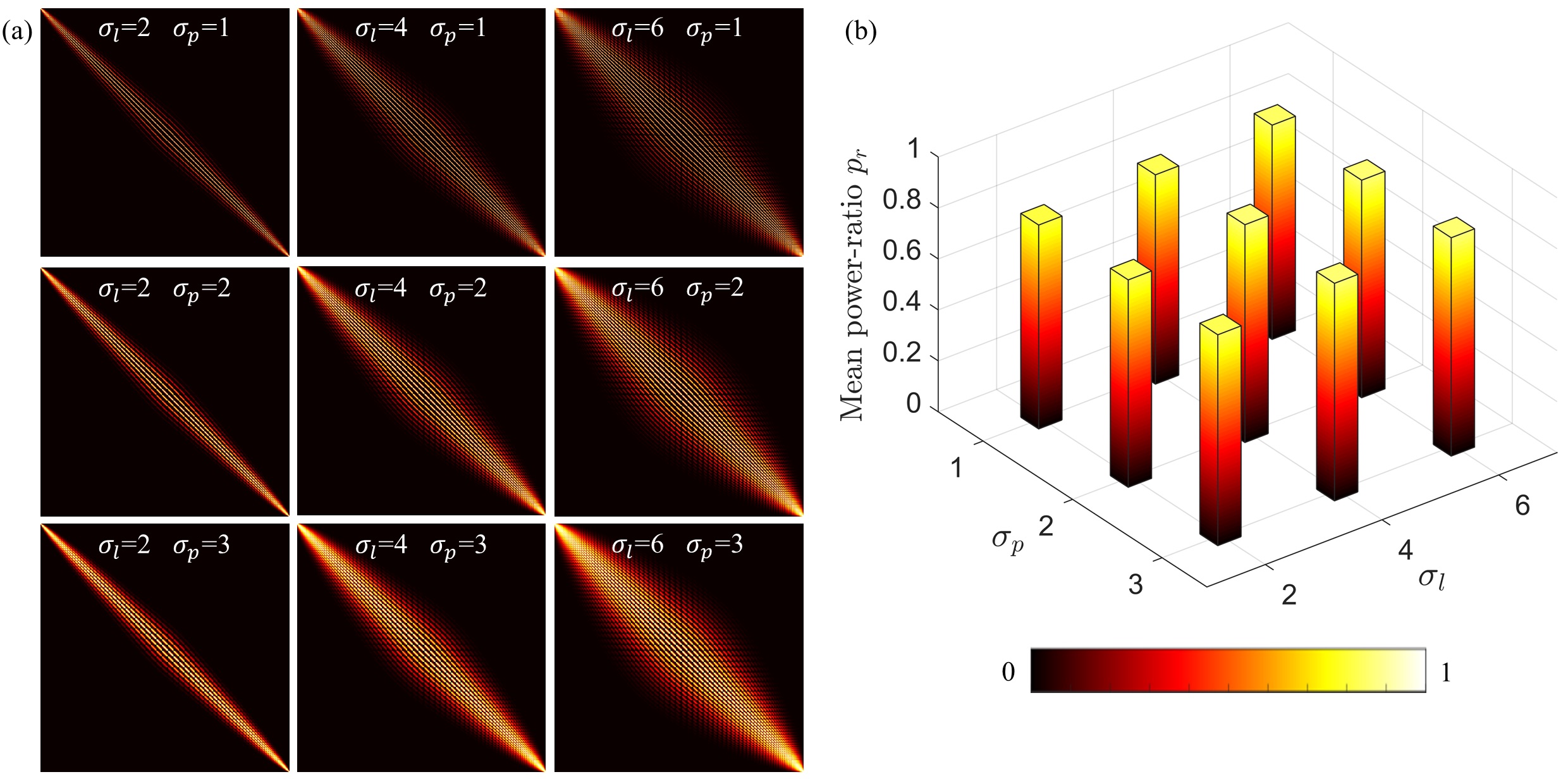}
\caption{{\bf Reconstruction with variation in the estimated support}: Here we reconstruct a TM with a compression ratio of $c=0.15$ and investigate the variation in reconstruction fidelity when the parameters of the estimated support are varied. {\bf(a)} Nine different estimated supports, with $\sigma_\ell = 2:2:6$, and $\sigma_p = 1:1:3$. {\bf(b)} Mean power-ratio of foci generated at the output using TMs reconstructed with the nine different supports shown in (a). In this case we find that even when the extent of the anticipated off-diagonal power coupling in the TM is over-estimated or under-estimated, the fidelity of the reconstructed TM is always higher than without using a support: $p_r>0.8$ in every case tested here. This illustrates that the FISTA reconstruction is robust to inaccurate estimates of the support within the demonstrated range.}
\label{Fig:support}
\end{figure*}

\renewcommand{\thefigure}{S6}
\begin{figure*}[h]
\includegraphics[width=10cm]{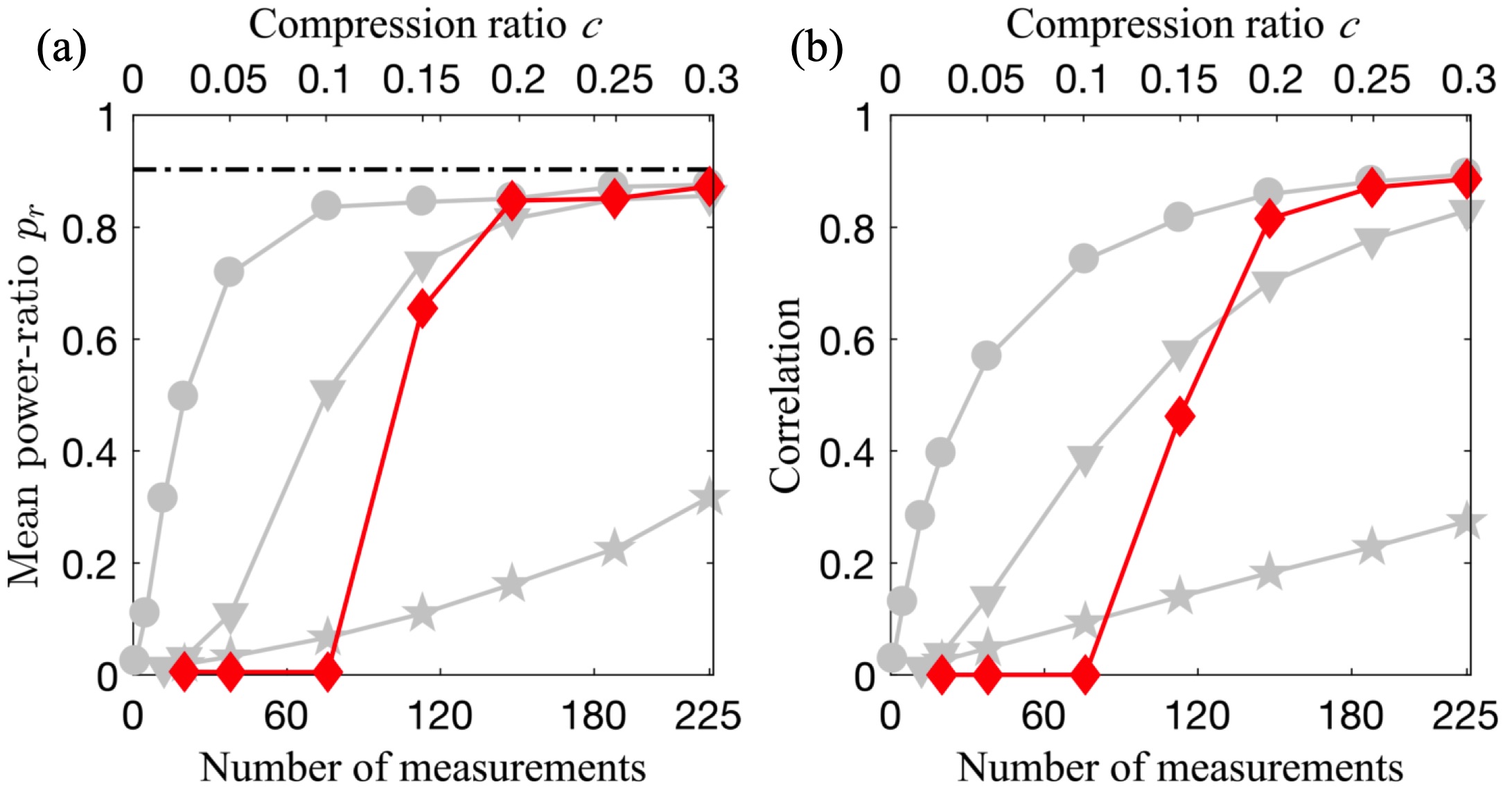}
\caption[TM reconstruction using Tikhonov regularisation]{{\bf TM reconstruction using Tikhonov regularisation}: In addition to the three reconstruction methods presented in the main paper, we also investigated a reconstruction method based on Tikhonov regularisation. In this case the estimated TM amplitude support was thresholded to create a binary mask: containing 0 in regions in which we expect there to be minimal power in the reconstructed TM, and 1 otherwise. To promote solutions that have low absolute values in regions specified by the predicted mask, the information about which regions of the TM we expect to be zero can be inserted into Eqn.~\ref{Eqn:sensing} by adding extra rows to $\VecS$ and $\Vecy$. For example, to specify that the $n^{\rm{th}}$ entry in $\Vect$ is 0, we vertically concatenate $\VecS$ with an extra row consisting of all elements set to 0 except for the $n^{\rm{th}}$ element which is set to 1. We also vertically concatenate $\Vecy$ with an extra element set to 0. This can be repeated for every element of the TM that we expect to be 0 according to the predicted binary mask. Note that the memory requirements are low for this approach as the extra rows are mainly zeros and can be represented using sparse matrices.
As long as enough additional information has been inserted into Eqn.~\ref{Eqn:sensing} to render it full-rank, then Eqn.~\ref{Eqn:sensing} can be solved using standard fast methods that minimise an error term $\eta$ given by the square of the Euclidean norm of the residual: $\eta = \|\VecS \Vect -  \Vecy \|_2^2$, which attempts to account for any inconsistencies in Eqn.~\ref{Eqn:sensing} due to inaccuracies in the estimation of the support or noise in the measurements. Additionally, the strength of the predicted support priors can be weighted with respect to the probe measurements by a factor $\lambda_{\rm Tik}$, for example, using the methods described in the SI of ref.~\cite{phillips2017adaptive}.
This method is a form of Tikhonov regularisation: to demonstrate this, let $\boldsymbol{\Gamma}$ represent the matrix formed from the extra rows vertically concatenated with $\VecS$ as described above, and $\boldsymbol{0}$ is a column vector representing the extra rows of zeros vertically concatenated with $\Vecy$. Therefore the new matrix-vector equation becomes: 
$\left[ \begin{array}{c} \VecS \\ \lambda_{\rm Tik}^{{1}/{2}} \boldsymbol{\Gamma} \end{array} \right] \Vect = \left[ \begin{array}{c} \Vecy \\ \boldsymbol{0} \end{array} \right] $,
where ${\lambda_{\rm Tik}^{1/2}}$ is the weighting factor. In this case the square of the Euclidean norm of the residual is given by $\left\lVert \left[ \begin{array}{c} \VecS \\ \lambda_{\rm Tik}^{1{/}2} \boldsymbol{\Gamma} \end{array} \right]\Vect -  \left[\begin{array}{c} \Vecy \\ \boldsymbol{0} \end{array} \right]\right\rVert_2^2 =
\left\lVert\left[ \begin{array}{c} \VecS\Vect - \Vecy \\ \lambda_{\rm Tik}^{{1}/{2}} \boldsymbol{\Gamma} \Vect \end{array} \right] \right\rVert_2^2 = \left\lVert\VecS\Vect-\Vecy\right\rVert_2^2 + \lambda_{\rm Tik}\left\lVert\boldsymbol{\Gamma}\Vect\right\rVert_2^2$, which is equivalent to Tikhonov regularisation with a Tikhonov matrix $\boldsymbol{\Gamma}$ and weighting factor $\lambda_{\rm Tik}$. Here we set $\lambda_{\rm Tik} = 1$.
{\bf{(a)}} and {\bf{(b)}} show the performance of Tikhonov regularisation (diamonds) in comparison with the other three reconstruction strategies (data shown in grey equivalent that in Figs.~\ref{Fig:compression}(c,d)). In this case for compression ratios of $c>0.25$, the performance of Tikhonov regularisation is equivalent to FISTA using sparsity and support priors, but the solution is returned approximately an order of magnitude faster. However, the solution is sensitive to under-estimations of the level of off-diagonal power spread in the TM. Additionally, for lower compression ratios of $c<0.2$, Eqn.~\ref{Eqn:sensing} is not full-rank, and since no sparsity priors are invoked the Tikhonov reconstruction undergoes catastrophic failure. In comparison, all other reconstruction strategies undergo graceful failure as $c$ is reduced. Therefore, although relatively fast to perform, Tikhonov reconstruction can only be successfully applied for mild compression ratios in circumstances where the level of off-diagonal power spread can be safely over-estimated.}
\label{Fig:Tikhonov}
\end{figure*}

\renewcommand{\thefigure}{S7}
\begin{figure*}[h]
\includegraphics[width=15cm]{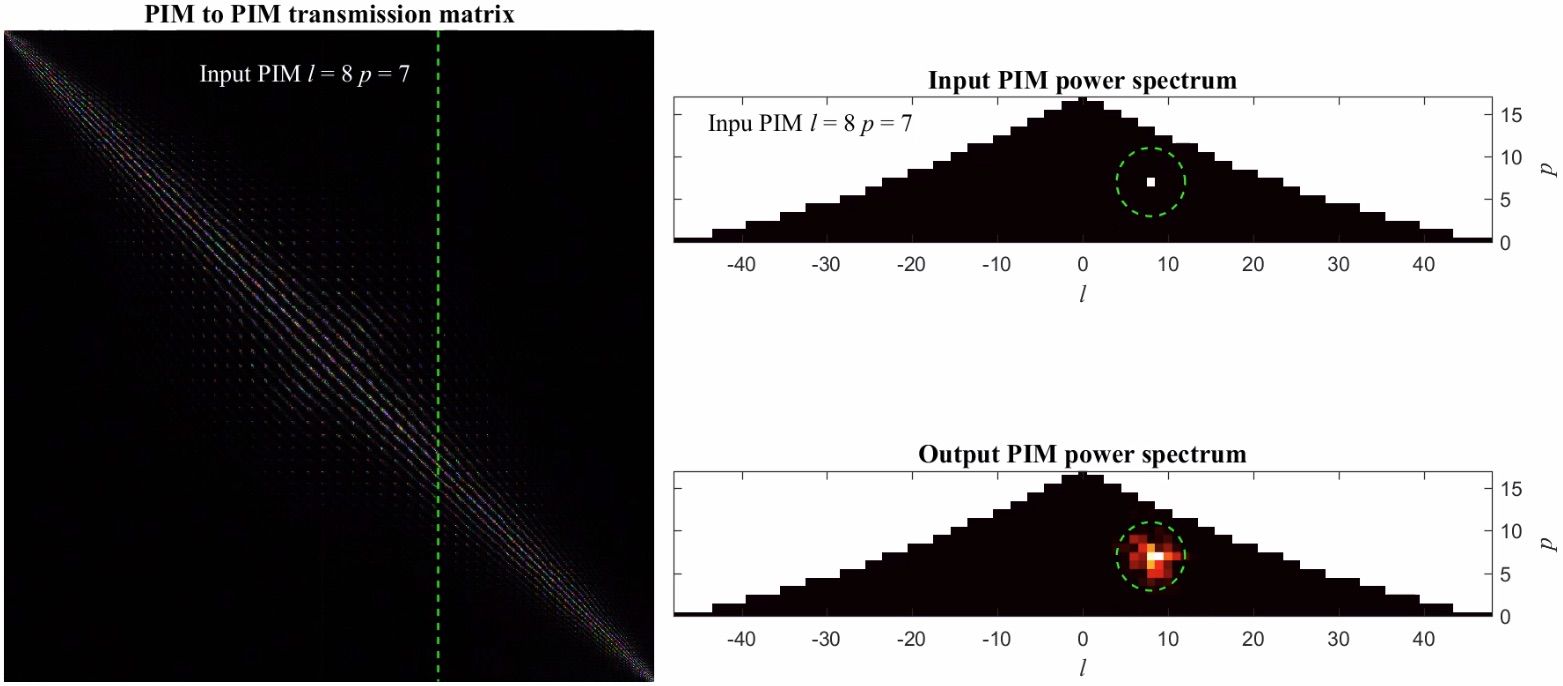}
\caption{{\bf Movie 1: Experimentally measured coupling of PIMs}: The figure depicts a frame from supplementary movie 1 showing the level of coupling each input PIM undergoes due to propagation through the MMF. The left hand panel shows the measured TM. The vertical green line highlights a single TM column. The n$^{\mathrm{th}}$ column corresponds to the n$^{\mathrm{th}}$ input mode. Each row captures the degree of coupling with all other PIMs experienced by the n$^{\mathrm{th}}$ input PIM after propagating through the MMF. The upper right hand plot highlights the n$^{\mathrm{th}}$ input PIM in ($\ell$,$p$)-space with a bright point. The lower right hand plot shows the level of power coupling of the n$^{\mathrm{th}}$ PIM at the fibre output. We see that power couples locally in ($\ell$,$p$)-space. We exploit this to predict the support of the TM. We also note that PIMs with low $p$-indices tend to couple more broadly to the neighbouring PIMs, as the values of the phase velocities of these PIMs are closer. There is potential for this observation to be exploited as more detailed prior knowledge about the structure of the TM in the future.}
\label{Fig:support}
\end{figure*}

\end{document}